\documentclass[twoside,english,authoryear]{elsarticle}
\usepackage[T1]{fontenc}
\usepackage[latin9]{inputenc}
\usepackage[a4paper]{geometry}
\usepackage{units}
\usepackage{textcomp}
\usepackage{amsmath}
\usepackage{amssymb}
\usepackage{graphicx}
\usepackage{setspace}
\usepackage{subscript}
\makeatletter
\journal{Planetary and Space Science}

\usepackage{amssymb}
\usepackage{pifont}

\makeatother

\begin{document}

\title{Global impacts of a Foreshock Bubble: Magnetosheath, magnetopause
and ground-based observations}

\author[imp]{M.~O.~Archer\fnref{fn1}\corref{cor1}}

\ead{m.archer10@imperial.ac.uk}

\author[ucla]{D.~L.~Turner}

\author[imp]{J.~P.~Eastwood }

\author[imp]{S.~J.~Schwartz }

\author[imp]{T.~S.~Horbury }

\cortext[cor1]{Corresponding author}

\address[imp]{Space \& Atmospheric Physics Group, The Blackett Laboratory, Imperial
College London, Prince Consort Road, London, SW7 2AZ, UK.}

\address[ucla]{Department of Earth, Planetary and Space Sciences, University of
California, 603 Charles E. Young Drive, Los Angeles, California, CA
90095-1567, USA.}
\begin{abstract}
Using multipoint observations we show, for the first time, that Foreshock
Bubbles (FBs) have a global impact on Earth's magnetosphere. We show
that an FB, a transient kinetic phenomenon due to the interaction
of backstreaming suprathermal ions with a discontinuity, modifies
the total pressure upstream of the bow shock showing a decrease within
the FB's core and sheath regions. Magnetosheath plasma is accelerated
towards the the intersection of the FB's current sheet with the bow
shock resulting in fast, sunward, flows as well as outward motion
of the magnetopause. Ground-based magnetometers also show signatures
of this magnetopause motion simultaneously across at least 7 hours
of magnetic local time, corresponding to a distance of 21.5~R\textsubscript{E}
transverse to the Sun-Earth line along the magnetopause. These observed
global impacts of the FB are in agreement with previous simulations
and in stark contrast to the known localised, smaller scale effects
of Hot Flow Anomalies (HFAs).\end{abstract}
\begin{keyword}
Magnetosphere \sep Foreshock \sep Transient \sep Dynamics \sep
Pressure
\end{keyword}
\maketitle

\section{Introduction\label{sec:Introduction}}

Although Earth's bow shock primarily mediates the solar wind flow
forming the magnetosheath, it is also an effective accelerator of
energetic particles allowing a portion of those incident to travel
back upstream along magnetic field lines forming Earth's foreshock
\citep[e.g. the review of][]{eastwood05}. The suprathermal backstreaming
particles in this region, which is typically spatially extended upstream
of the quasi-parallel shock (where the acute shock normal - magnetic
field angle $\theta_{Bn}\lesssim$45\textdegree{}), cause kinetic
instabilities within the incident solar wind plasma that can generate
ultra-low frequency (ULF) waves \citep[e.g.][]{hoppe81} and in turn
scatter particles. The foreshock is highly dynamic, due to variations
in the interplanetary magnetic field (IMF) and solar wind conditions,
and a number of kinetic phenomena resulting from the interaction of
such changes with the quasi-parallel bow shock have been both simulated
and observed. These foreshock transients, which include hot flow anomalies
\citep{schwartz85}, foreshock cavities \citep{thomas88} and the
recently discovered foreshock bubbles \citep{omidi10}, can have significant
magnetospheric impacts such as perturbing the magnetopause \citep{sibeck99,turner11}
and generating magnetospheric ULF waves \citep{fairfield90,eastwood11,hartinger13}.

\begin{figure}
\begin{centering}
\includegraphics{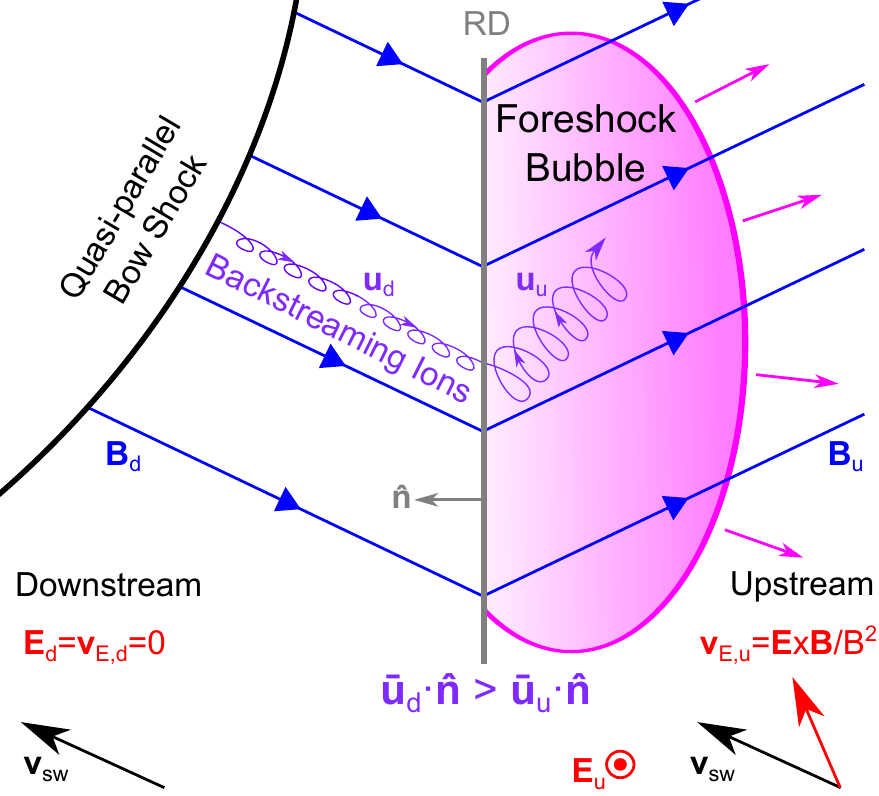}
\par\end{centering}

\centering{}\caption{Example schematic of Foreshock Bubble formation. A rotational disconituity
(RD, grey) which increases the angle between the IMF (blue) and the
Sun-Earth line on its upstream side results in an greater upstream
motional electric field $\mathbf{E}=-\mathbf{v}_{sw}\times\mathbf{B}$
(red, out of the page). The motion of backstreaming ions in the foreshock
on the downstream side of the RD is altered upstream, with a larger
guiding centre drift $\mathbf{v}_{E}$ (red arrow) back towards the
RD as well as increased pitch angle due to the IMF change, resulting
in an increase in the suprathermal density and temperature upstream
of the RD. This increase in thermal pressure causes a local expansion,
forming a Foreshock Bubble.\label{fig:FB-cartoon}}
\end{figure}

Foreshock Bubbles (FBs), first predicted by 2D kinetic hybrid simulations
\citep{omidi10,omidi13,karimabadi14}, are transient phenomena caused
by the interaction of suprathermal backstreaming ions with a (rotational)
discontinuity. Figure \ref{fig:FB-cartoon} shows an example schematic
of how FBs are thought to form, following \citet{turner13}. The motion
of backstreaming ions, moving along the magnetic field and originating
from the quasi-parallel bow shock, may be altered upon encountering
a rotational discontinuity (RD). If the IMF cone angle $\theta_{Bx}$
(the angle between the IMF and the Sun-Earth line) is increased on
the upstream side of this discontinuity, then the motional electric
field $\mathbf{E}=-\mathbf{v}_{sw}\times\mathbf{B}$ will be greater
and the backstreaming particles will experience increased $\mathbf{E}\times\mathbf{B}$
guiding centre drift $\mathbf{v}_{E}$ equal to the component of the
solar wind velocity perpendicular to the magnetic field \citep{greenstadt76}
i.e. with a component back towards the RD. In addition, the IMF change
also results in the backstreaming ions' pitch angles increasing thereby
converting some of the ions' motion parallel to the magnetic field
into gyromotion. It can be shown (see \ref{sec:Suprathermal-flux-derivation})
in the deHoffmann-Teller rest frame of the RD \citep{dehoffman50},
where the motional electric field is zero on both sides and thus particle
energies are conserved, that the increase in particle pitch angle
results in a concentration of suprathermal ion density upstream of
the RD. Together with the increase in gyrospeed, the temperature and
thermal pressure of the plasma increase upstream of the RD, thereby
causing the thermal plasma to expand. Due to this expansion against
the solar wind, a hot core region of depleted density and magnetic
field with significant flow deflections forms immediately upstream
of the RD followed by a compressed ``sheath'' region and possibly
a shock. This whole structure, which convects with the RD whilst also
growing, is what is known as a Foreshock Bubble.

The signatures of an FB in spacecraft observations, however, exhibit
many similarities with Hot Flow Anomalies (HFAs): a transient phenomenon
in the vicinity of the intersection of the bow shock with a (tangential)
discontinuity due to kinetic shock processes \citep{schwartz85,schwartz88,thomsen88,paschmann88}.
An HFA consists of a hot depleted core, usually on the side of the
current sheet with quasi-parallel bow shock conditions \citep{omidi07,zhang10,wang13b},
sandwiched by compressions and sometimes shocks on both sides due
to the lateral expansion of the plasma \citep{fuselier87,lucek04b}.
This structure tracks across the bow shock with a transit velocity
given by \citep{schwartz00}
\begin{equation}
\mathbf{v}_{trans}=\frac{\mathbf{v}_{sw}\cdot\mathbf{n}_{DD}}{\sin^{2}\theta_{bs,DD}}\left(\mathbf{n}_{DD}-\cos\theta_{bs,DD}\mathbf{n}_{bs}\right)
\end{equation}
where $\mathbf{n}_{bs}$ and $\mathbf{n}_{DD}$ are the normals to
the bow shock and directional discontinuity (DD) respectively, $\theta_{bs,DD}$
is the angle between these and $\mathbf{v}_{sw}$ is the solar wind
velocity. \citet{schwartz00} summarised a set of conditions for the
formation of HFAs, which required that the motional electric field
points into the discontinuity on at least one side and that the transit
speed of the discontinuity $v_{trans}$ is much slower than the gyrospeed
of ions reflected at the bow shock. Furthermore, they showed that
HFAs preferentially occur if the discontinuity is tangential in nature
(with no magnetic flux threading the current sheet), exhibits a small
jump in magnetic field strength and quasi-perpendicular bow shock
conditions are present on at least one side (with the upstream/trailing
edge being favourable).

\citet{turner13} presented the first observational evidence of FBs
upstream of Earth's bow shock, comparing and constrasting their signatures
to HFAs. They developed a set of identification criteria to distinguish
between the two phenomena:
\begin{enumerate}
\item HFA formation requires the discontinuity intersects with the bow shock;
FB formation does not.
\item HFA cores form on the quasi-parallel side of the discontinuity or
centred on the discontinuity if perpendicular/parallel on both sides
\citep{omidi07,zhang10,wang13b}; FB cores should only form upstream
of the discontinuity.
\item HFAs tend to be bounded on both sides by compression regions except
theoretically when the ratio of incident suprathermal to solar wind
ions $\gtrsim65\%$ \citep{thomsen88}, though the strength of the
compressions is often asymmetric with the upstream one typically being
much larger \citep[e.g.][]{paschmann88}; FBs observed from within
the foreshock should be bounded by a compression region or shock on
the upstream side only.
\item HFAs require the electric field point into the discontinuity on at
least one side; FBs do not.
\item HFA boundaries can exhibit a range of orientations \citep{paschmann88}
though are often close to that of the discontinuity due to the lateral
expansion of plasma; FB boundary normals observed from within the
foreshock should be oriented predominantly sunwards.
\item HFAs move along the bow shock with the discontinuity intersection;
FBs should convect with the solar wind.
\item HFAs have transverse sizes up to $\sim$4~R\textsubscript{E} \citep{facsko09}
and their features are thought to diminish within $\sim$5~R\textsubscript{E}
of the bow shock \citep{wang13a,archer14}; FBs might have transverse
scales comparable with the size of the quasi-parallel bow shock, $\sim$10~R\textsubscript{E}
or more \citep{omidi10}.
\end{enumerate}
HFAs are known to have fairly localised impacts which track across
the magnetosphere, including flow deflections in the magnetosheath,
distortions of the magnetopause over $\sim$5~R\textsubscript{E},
and travelling convection vortices in the ionosphere \citep{sibeck99,eastwood08,jacobsen09,archer14}.
In contrast, the impacts of FBs are predicted by simulations to be
global in scale \citep{omidi10}: the arrival of the structure at
the bow shock causes reversal of the magnetosheath flow back towards
the FB core due to its reduced pressure compared to the magnetosheath
plasma, in turn resulting in large scale outward motion of the magnetopause.
\citet{hartinger13} presented observations of the magnetospheric
response to an FB at a single spacecraft location, consisting of a
rarefaction (due to the reduced dynamic pressure of the FB core) and
then compression (due to the enhanced dynamic pressure of the FB sheath
and shock) of the magnetospheric magnetic field and accompanied by
Pc5 (2-7~mHz) ULF wave activity in the perpendicular components.
However, the scale size of the magnetospheric impact of FBs has yet
to be determined observationally. Since Pc5 ULF waves play a role
in the mass, energy and momentum transport within the Earth\textquoteright{}s
magnetosphere e.g. accelerating electrons in auroral regions \citep{lotko98}
and the radiation belts \citep{claudepierre13,mann13}, it is important
to understand the impacts of drivers of magnetospheric dynamics such
as FBs. In this paper we present observationally, for the first time,
the response of the magnetosheath and magnetopause to an FB, using
multipoint spacecraft observations in conjunction with ground magnetometer
measurements. We demonstrate the global nature of the transient's
impact, in agreement with the suggestion of previous simulations and
in stark contrast to the known localised effects of HFAs.

\section{Observations}

\subsection{Solar Wind \& Foreshock}

\begin{figure}
\begin{centering}
\includegraphics{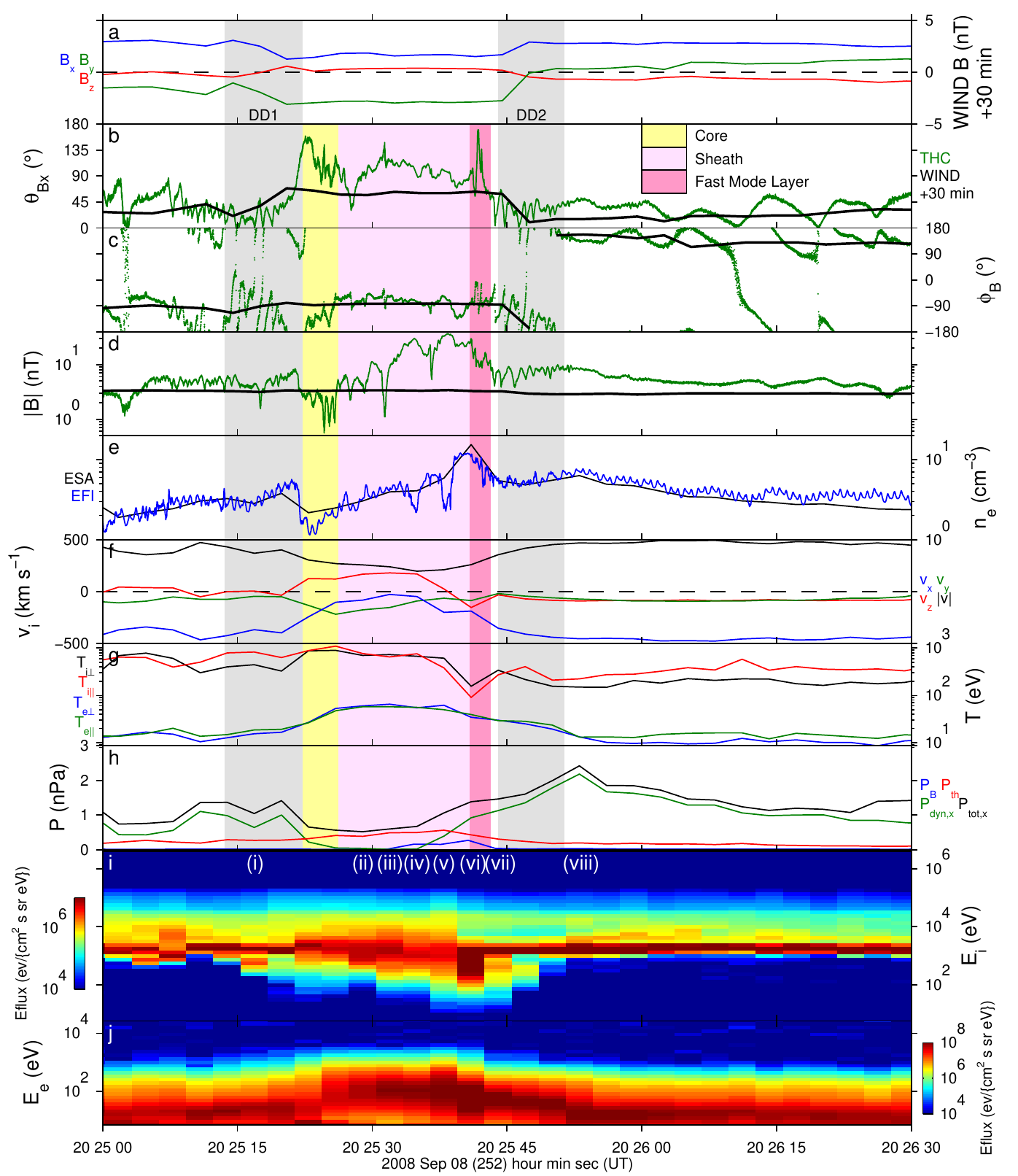}
\par\end{centering}

\caption{THC observations $\sim$1 R\textsubscript{E} upstream of Earth's
bow shock. From top to bottom: GSE magnetic field components in the
pristine solar wind observed by WIND (xyz in blue, green, red) near
L1 where data has been lagged by 30 min; comparisons between WIND
(black) and THC (green) of the cone angle $\theta_{Bx}$ (between
the magnetic field and the GSE x-direction), GSE clock angle $\phi_{B}$,
and magnetic field strength; electron density from ESA (black) and
EFI (blue); GSE ion velocity components (xyz in blue, green, red)
and magnitude (black); ion and electron temperatures parallel (red
for ions, green for electrons) and perpendicular (black for ions,
blue for electrons) to the magnetic field; magnetic (blue), thermal
(red), anti-sunward dynamic (green) and total anti-sunward (black)
pressures; ion and electron energy spectrograms where the colour scale
represents the differential energy flux. Two directional discontinuities
(DD1 \& DD2) are highlighted in grey and core (yellow), sheath (light
pink) and fast mode layer (pink) regions are also indicated.\label{fig:THC-tseries}}
\end{figure}

On 08 September 2008 at around 20:25 UT, two of the THEMIS \citep{angelopoulos08}
spacecraft were located in Earth's foreshock at 14.3~R\textsubscript{E}
(THC) and 16.0~R\textsubscript{E} (THB) upstream of the Earth. Figure
\ref{fig:THC-tseries}b-d show high resolution (128~vectors~s\textsuperscript{-1})
Fluxgate Magnetometer (FGM) data at THC in green. It is clear in the
angles $\theta_{Bx}$ (the cone angle between the IMF and the GSE
x-direction) and $\phi_{B}$ (the GSE clock angle) that there were
two directional discontinuities (DDs) which changed the IMF's orientation.
We denote these as DD1 and DD2 and highlight them as the grey areas
in Figure \ref{fig:THC-tseries}. The same directional discontinuities
were also observed earlier by both ACE's Magnetic Field Experiment
\citep{smith98} (not shown) and WIND's Magnetic Field Investigation
\citep{lepping95} in the pristine solar wind near L1, with 30~min
lagged data from the latter shown at 3~s resolution in black, revealing
good agreement with the THC observations. Due to the proximity of
the spacecraft to the subsolar bow shock, the IMF - shock normal angle
$\theta_{Bn}$ magnetically connected to both THEMIS spacecraft was
very similar to $\theta_{Bx}$ i.e. the bow shock was quasi-parallel
before DD1, turning quasi-perpendicular between the two current sheets
and subsequently returning to quasi-parallel conditions again. While
the magnitude of the IMF as observed by WIND was relatively steady
(panel d), THC observed strong variations in magnetic field strength
following DD1: firstly decreasing sharply (yellow area), then increasing
over $\sim$12~s (light pink area), subsequently sharply dropping
(dark pink area) before DD2 was observed and eventually returning
to the ambient value after DD2.

\begin{figure}
\begin{centering}
\includegraphics{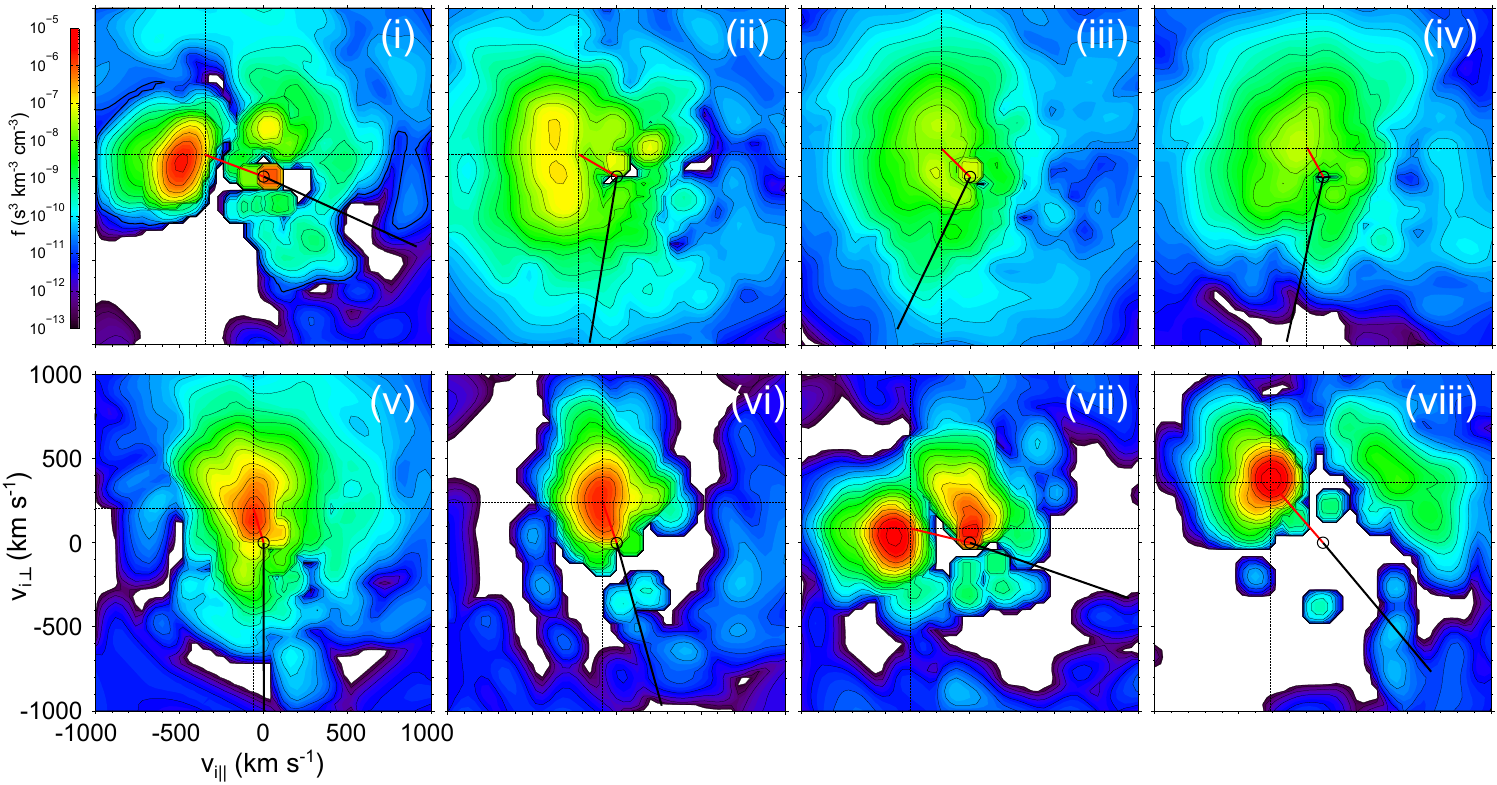}
\par\end{centering}

\caption{2D cuts of ion distribution functions in the B-v plane at THC corresponding
to the times indicated in Figure \ref{fig:THC-tseries}. The black
circle denotes the origin in the spacecraft frame, the red line is
the velocity moment and the black line shows the projection of the
GSE x direction.\label{fig:THC-distributions}}
\end{figure}

Plasma data from ACE's Solar Wind Electron Proton Alpha Monitor \citep{mccomas98}
and WIND's Solar Wind Experiment \citep{ogilvie95} revealed no strong
variations (not shown due to their low time resolutions of 60~s and
97~s respectively). We present combined ``burst'' mode Electrostatic
Analyzer (ESA) \citep{mcfadden08a} and Solid State Telescope (SST)
data (3~s cadence) at THC in Figure \ref{fig:THC-tseries}e-j, with
corresponding 2D cuts of the ion distribution functions in the B-v
plane shown in Figure \ref{fig:THC-distributions}. We also estimate
the electron density from the 128~samples~s\textsuperscript{-1}
spacecraft potential Electric Field Instrument \citep{bonnell08}
data using the method described by \citet{chen12} (shown in blue
in panel e), which clearly reveal similar variations to $\left|\mathbf{B}\right|$.
Large deflections of the ion velocity were observed (panel f) during
this event, predominantly because the x component almost went to zero,
reducing the solar wind speed by about half. The electron temperatures
parallel (green) and perpendicular (blue) to the magnetic field both
increased inside the event (panel g). Figure \ref{fig:THC-distributions}i
shows that before the event the ion distribution consisted of the
solar wind beam along with intermediate backstreaming suprathermal
ions characteristic of Earth's foreshock \citep[e.g.][]{fuselier95}.
This backstreaming population results in the large parallel (red)
temperature moment (over the entire distribution) in Figure \ref{fig:THC-tseries}g.
In the depleted core region (yellow area in Figure \ref{fig:THC-tseries})
there was clearly an increase in the ion temperature, which also became
more isotropic, and these hot, isotropic ion temperatures remained
throughout the compressed ``sheath'' region (light pink), with the
distributions (Figure \ref{fig:THC-distributions}ii-v) evolving to
a single component plasma. Following the ``sheath'', a correlated
decrease in the density and magnetic field was observed (dark pink
area in Figure \ref{fig:THC-tseries}) across which the velocity increased,
returning to almost the solar wind speed. While such signatures are
similar to that found exiting a shock transition into the unshocked
upstream solar wind, the electron temperatures decreased smoothly
rather than showing a sharp transition, and the ion temperature within
this layer is actually lower than the upstream solar wind values.
Due to these dissimilarities with shocks such as Earth's bow shock,
we shall refer to this transition simply as a fast mode layer (FML).
Immediately upstream of the FML (Figure \ref{fig:THC-distributions}vii)
the solar wind beam was observed once again along with a strong field
aligned beam (comparable in phase space density to the solar wind),
whereas following DD2 (Figure \ref{fig:THC-distributions}viii) an
intermediate ion distribution was observed.

We calculate the combined isotropic ion and electron thermal pressure
$P_{th}$ (red), the magnetic pressure $P_{B}$ (blue) and the anti-sunward
dynamic pressure $P_{dyn,x}$ (green) as well as the sum of these,
the total pressure $P_{tot,x}$ (black), with the results shown in
Figure \ref{fig:THC-tseries}h. During the event the total pressure
upstream of the bow shock was decreased from $\sim$1.2~nPa to 0.5~nPa,
principally because of the reduced dynamic pressure associated with
the velocity deflections. A total pressure increase to $\sim$2.4~nPa
was also observed due to a slight density increase after DD2. While
this might have been associated solely with DD2, it is possibly related
to the transient event associated with DD1 since the FML (before DD2)
did not return the density, magnetic field strength or electron temperatures
to their ambient solar wind values.

We note that THB, less than 2~R\textsubscript{E} further upstream
than THC and separated transversely by $\sim$1~R\textsubscript{E}
in the y and 2.5~R\textsubscript{E} in the z GSE directions, did
not encounter strong variations between DD1 and DD2 when observed
52~s earlier (not shown). $\left|\mathbf{B}\right|$ decreased by
$\sim$2~nT, the density dropped by $\sim$1.6~cm\textsuperscript{-3},
no compressions were observed, the velocity was deflected by only
$\sim$50~km~s\textsuperscript{-1} and ion temperature increases
were $\sim$100~eV (though in ``reduced'' mode as used here, ESA
does not fully resolve the solar wind beam \citep{mcfadden08a}).

\subsubsection{Analysis\label{sub:SW-Analysis}}

Despite the proximity of THC to the bow shock ($\sim$1.3~R\textsubscript{E}
upstream using the \citet{shue98} magnetopause model and the \citet{farrisrussell94}
bow shock standoff relation), the observed event cannot be explained
as being due to a breathing motion of the bow shock passing over the
spacecraft: the density did not increase as quickly as the magnetic
field in the ``sheath'' region \citep[e.g.][]{schwartz11}, the
greatest temperatures were observed in the region of decreased density
and magnetic field, and the velocity contained strong transverse deflections
inconsistent with a subsolar bow shock/magnetosheath encounter. This
event was therefore a foreshock transient \citep[e.g.][]{fairfield90}
associated with DD1 and here we perform further analysis in order
to aid its classification.

To estimate the orientation of DD1 we employ three spacecraft timing
\citep{horbury01b} using ACE, WIND and THC. We choose this method
rather than four spacecraft triangulation \citep[e.g.][]{schwartz98}
since the two THEMIS spacecraft were much closer together than those
at L1. This computed normal $\mathbf{n}_{DD1}=$ (0.54,0.56,0.63)
in GSE coordinates, resulted in agreement with the observed time lag
between THB and THC to within $\sim$5\textdegree{} of this orientation.
The fraction of magnetic flux threading the current sheet $\mathbf{B\cdot}\mathbf{n}_{DD1}/\left|\mathbf{B}\right|\sim$0.2
and the negligible change in $\left|\mathbf{B}\right|$ observed in
the pristine solar wind mean that DD1 was an ED (either a tangential
or rotational discontinuity) according to the classification of \citet{neugebauer84}.
The vast majority of solar wind current sheets are in this category
\citep[e.g.][]{knetter04}.

To estimate the normal to the FML we borrow methods from fast-mode
shocks, namely magnetic coplanarity, velocity coplanarity (valid for
high Mach shocks) and the three mixed mode normals \citep{schwartz98}.
The resulting normals from all these methods were in excellent agreement
at $\mathbf{n}_{FML}=$(0.87,-0.31,-0.31) in GSE. From this normal
we estimated the speed of the layer using both the conservation of
mass \citep{schwartz98} and continuity of the tangential electric
field \citep[c.f.][]{smith88}. Both methods again were in agreement
resulting in a normal speed in the spacecraft frame $v_{FML}=-100$~km
s\textsuperscript{-1}. We calculate (using the speed and orientation
of the FML along with the solar wind speed) that the transient was
rapidly expanding in the x direction at $\sim$400 km s\textsuperscript{-1}.

\begin{figure}
\begin{centering}
\includegraphics{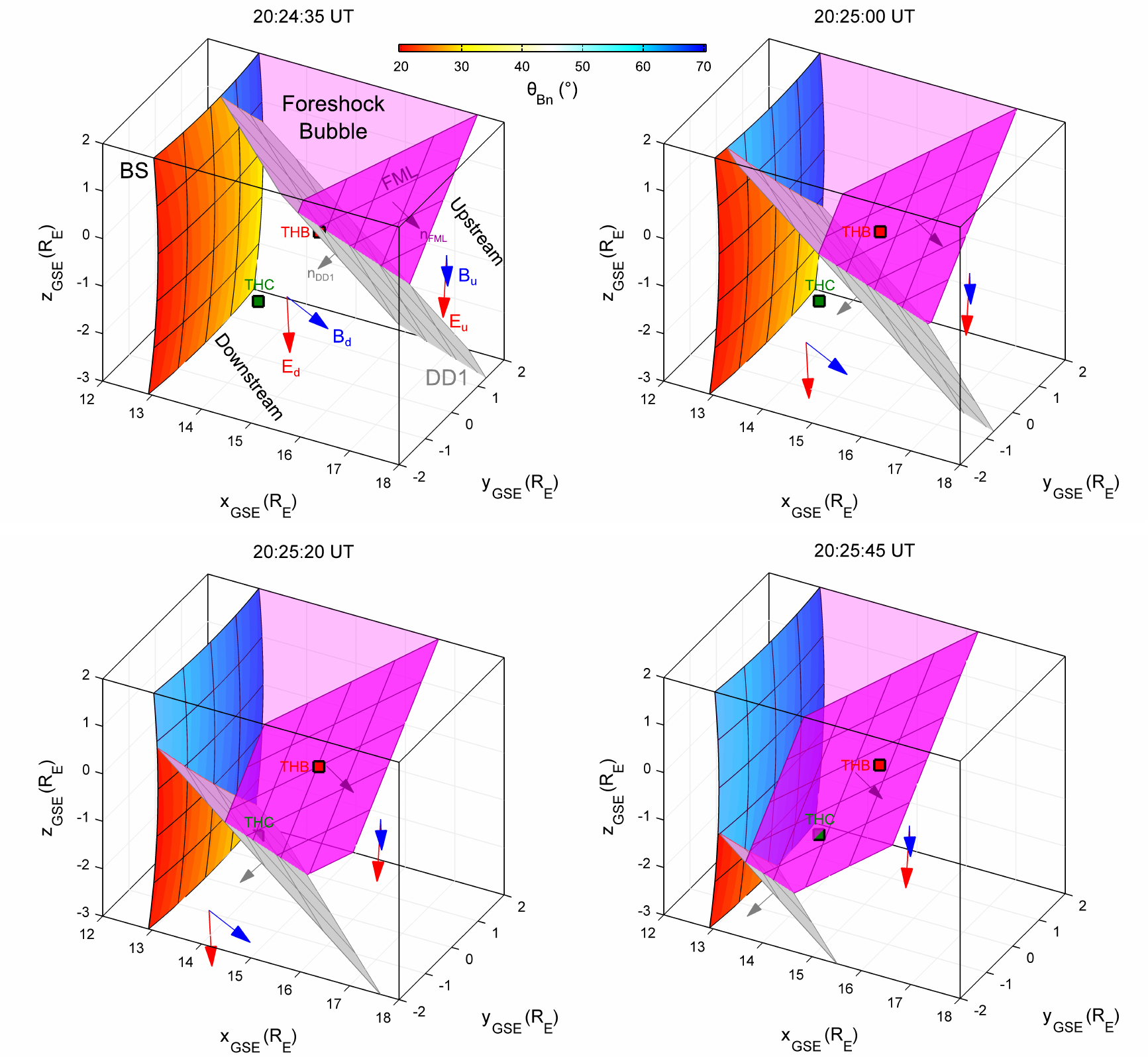}
\par\end{centering}

\caption{Schematics of the event in the solar wind/foreshock at different times.
Spacecraft positions shown as the coloured squares. The current sheet
DD1, with normal $\mathbf{n}_{DD1}$, and the fast mode layer (FML),
with normal $\mathbf{n}_{FML}$, are shown in grey and magenta respectively.
The \citet{farris91} model bow shock (BS) is coloured by the angle
$\theta_{Bn}$. The IMF (blue) and motional electric field (red) either
side of DD1 are displayed as arrows. \label{fig:event-schematic}}
\end{figure}

Given the calculated orientations and speeds of DD1 and the FML, it
is possible to construct 3D schematics of this event. These are shown
in Figure \ref{fig:event-schematic} at different times, assuming
that the FML was approximately planar over the spacecraft separations
and that its speed remained constant. The bottom left and right panels
shows when THC (green) observed DD1 and the FML respectively. In the
former, it is clear that THC was downstream of the FML at this time.
The top left panel corresponds to when THB (red) observed DD1, revealing
that the FML was just downstream of the spacecraft at this time, which
may explain why THB did not observe similar variations to THC.

\subsubsection{Classification}

The transient could have been a Hot Flow Anomaly \citep[e.g.][]{schwartz00}:
DD1 may have been tangential, it intersected the bow shock and the
motional electric field pointed towards it on the upstream side (Figure
\ref{fig:event-schematic}). However, the depleted core was located
on the side of the discontinuity with quasi-perpendicular bow shock
conditions, whereas HFA cores are displaced to the quasi-parallel
side \citep{omidi07,zhang10,wang13b}. If one assumes that the spacecraft
observations were due to the transient moving across the bow shock,
typical for HFAs \citep[e.g.][]{facsko09}, then due to the orientations
of $\mathbf{v}_{trans}$ and $\mathbf{n}_{FML}$ the observed core
and ``sheath'' regions must have been being upstream of the FML.
Such an arrangement is not consistent with the typical picture of
HFAs from simulations \citep{lin02} and observations \citep{lucek04b}.
Furthermore, DD1's fast transit speed of $0.7\times$ the reflected
ion gyrospeed \citep{schwartz00}, the lack of a significant compression
on the downstream side of the transient given that the suprathermal
ion density was 8\% of the solar wind density \citep{thomsen88},
and the predominantly sunward orientation of the FML normal do not
fit with the expected properties of an HFA at the subsolar bow shock.

Conversely, since DD1 may have been an RD and the IMF cone angle $\theta_{Bx}$
was increased on its upstream side, the event could have been a Foreshock
Bubble (c.f. Figure \ref{fig:FB-cartoon}). The location of the depleted
core and the compression on the upstream side of DD1 are consistent
with this, as is the sunward orientation of the FML \citep{turner13}.
The transient therefore satisfied all the FB identification criteria
of \citet{turner13} apart from the transverse size criterion, which
we are unable to test due to no additional spacecraft observations
of the transient. Henceforth we identify this event as a Foreshock
Bubble.

\subsection{Magnetosheath \& Magnetopause}

\begin{figure}
\begin{centering}
\includegraphics{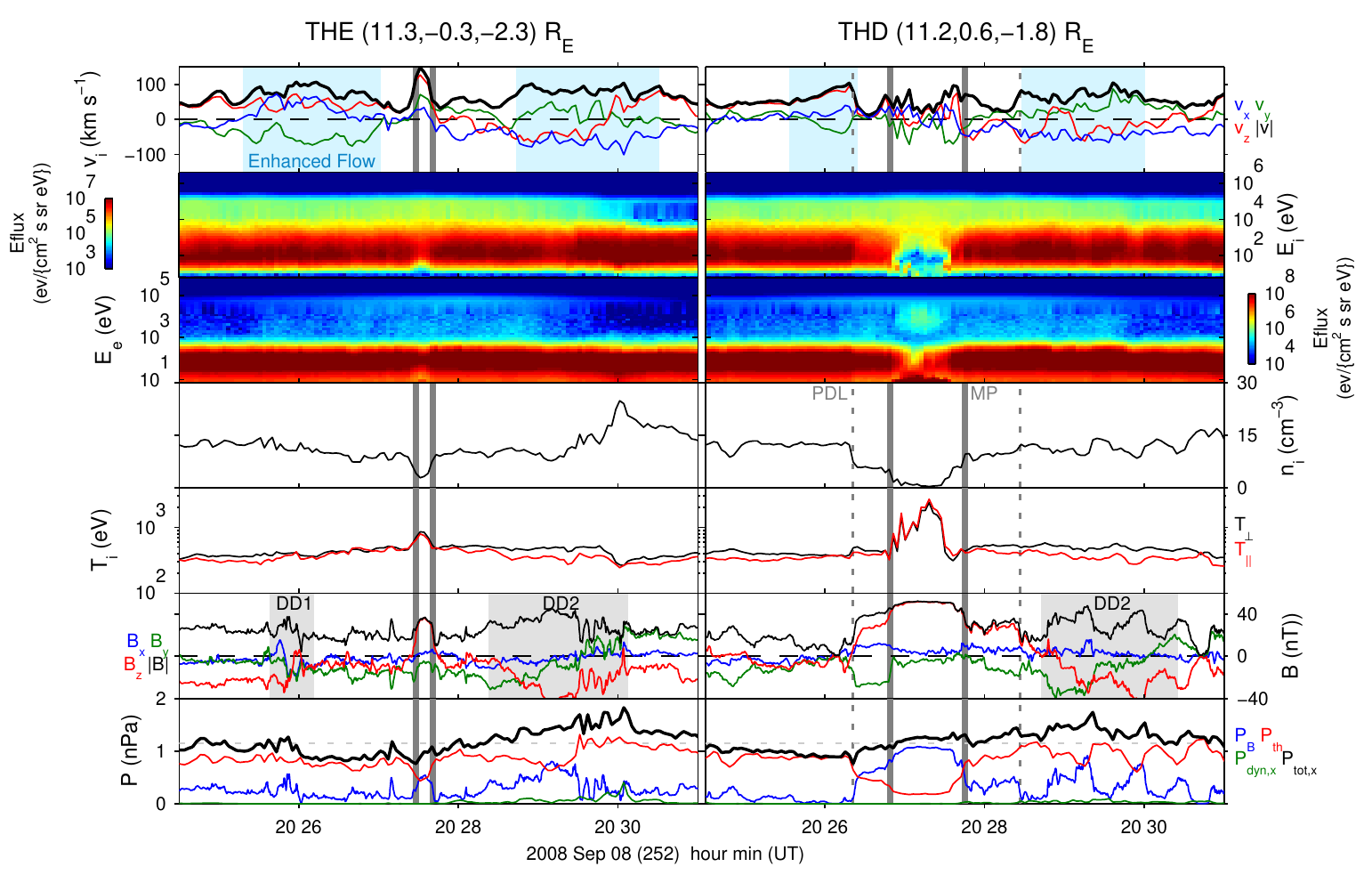}
\par\end{centering}

\caption{Magnetosheath observations from THE (left) and THD (right) with the
spacecraft positions in GSE noted. From top to bottom: the ion velocity
components in GSE (xyz in blue, green, red) and magnitude (black);
ion energy spectrogram where the colour scale represents the differential
energy flux; electron energy spectrogram; ion density; ion temperatures
parallel (red) and perpendicular (black) to the magnetic field; magnetic
field components in GSE (xyz in blue, green, red) and magnitude (black);
and the magnetic (blue), thermal (red), anti-sunward dynamic (green)
and total anti-sunward (black) pressures. The transmitted directional
discontinuities DD1 and DD2, identified in Figure \ref{fig:THC-tseries},
are highlighted in grey. Blue shaded regions indicate enhanced magnetosheath
flows, vertical solid grey lines correspond to magnetopause crossings
and the vertical dotted line shows the edge of the plasma depletion
layer.\label{fig:MSH}}
\end{figure}

During this interval, another two of the THEMIS spacecraft (THD \&
THE) were located in the magnetosheath $\sim$1 R\textsubscript{E}
sunward of the \citet{shue98} model magnetopause, with THE slightly
further away from the Earth by $\sim$0.2 R\textsubscript{E}, and
separated by $\sim$1 R\textsubscript{E} chiefly in the GSE y direction.
Figure \ref{fig:MSH} shows combined ``reduced'' mode ESA and SST
data (3~s resolution) as well as 4~vectors~s\textsuperscript{-1}
FGM data from the two spacecraft. In the latter we identify the transmitted
discontinuities DD1 and DD2 (highlighted again by grey areas) by comparing
the GSE y and z components with the WIND observations (Figure \ref{fig:THC-tseries}a).
While THE (left) observed both of these, only DD2 could be clearly
identified at THD (right). THE observed DD1 $\sim$35~s later than
it was observed at THC upstream of the bow shock, in fair agreement
with the expected 45~s lag from two spacecraft timing given the current
sheet's estimated orientation (making the reasonable assumption that
the magnetosheath did not significantly alter this \citep[c.f.][]{sibeck03}).
Approximately centred around the transmitted discontinuity, THE (left)
observed an enhancement of the magnetosheath flow speed (top panel,
shaded in blue) from $\sim$47 km s\textsuperscript{-1} to $\sim$107
km s\textsuperscript{-1} lasting about $1\nicefrac{3}{4}$~min.
This enhanced flow also exhibited large deflections such that the
usual anti-sunward magnetosheath was accelerated both sunwards and
in the GSE -y direction. At around the same time, a flow enhancement
was also observed at THD (right) with a similar magnitude. On the
other hand, the direction of the flow was only marginally sunward
and exhibited some acceleration in the +z direction as well as the
main deflection towards -y as with THE.

Again we calculate the thermal, magnetic, antisunward dynamic (for
only those intervals in which the flow was antisunward) and total
pressures in the magnetosheath as shown in the bottom panels of Figure
\ref{fig:MSH}. These reveal a total pressure decrease in the magnetosheath
of $\sim$0.3~nPa at 20:26, around the time of the enhanced flows.
Whereas at THD (right) the pressure decrease is fairly gradual, at
THE (left) this drop is sharp and follows the observation of the transmitted
discontinuity DD1. This suggests that it was in response to the reduced
total pressure associated with the FB that formed on the upstream
side of this current sheet (Figure \ref{fig:THC-tseries}h).

Following the enhanced flow and pressure decrease, the magnetopause
moved outwards causing THD (right) to enter into the plasma depletion
layer (vertical dotted line) \citep{zwan76} between 20:26:20-20:26:50
UT, followed by the magnetopause passing over the spacecraft (solid
vertical line). The spacecraft had a brief excursion in the magnetosphere
before encountering the boundary again and returning to the magnetosheath
at around 20:27:45 UT. In contrast, THE (slightly further from the
model boundary) only partially encountered the magnetopause and did
not fully cross the boundary into the magnetosphere.

We applied minimum variance analysis (MVA) \citep{sonnerup98} to
spin-resolution FGM data to determine normals for the observed magnetopause
crossings, testing the quality of the analysis via the intermediate-to-minimum
eigenvalue ratio test ($\lambda_{int}/\lambda_{min}\gtrsim10$ implying
a reliable normal) as well as the sensitivity of the resulting normal
to different time intervals centred on the boundary. Reliable normals
could only be obtained for the THD crossings, since MVA was poorly
conditioned at THE with $\lambda_{int}/\lambda_{min}\sim2$. For the
inbound case the obtained normal was deflected 18\textdegree{} from
the \citet{shue98} model boundary, predominantly in the GSE +y direction,
whereas for the outbound crossing the difference was only 6\textdegree{}
i.e. consistent with being zero within errors \citep{sonnerup71,khrabrov98}.
These normals therefore suggest global/large scale motion of the magnetopause.

We estimate the normal speed of the boundary using the two spacecraft
timing method \citep[e.g.][]{schwartz98}. We apply this only to the
inbound crossings, resulting in a speed of only $\sim$20 km s\textsuperscript{-1}.
While this method can be unreliable when the spacecraft separation
is predominantly perpendicular to the boundary normal, as is the case
here, the calculated speed is similar to the observed plasma velocity
ahead of the boundary. By the time the magnetopause was observed by
the spacecraft we deduce (both from the observed magnetosheath flow
and the timing) that it had substantially decelerated. The boundary
had moved outwards by $\sim$1 R\textsubscript{E} from its nominal
location, while a simple pressure balance calculation using the total
pressure within the FB observed by THC (Figure \ref{fig:THC-tseries}h)
results in an expected 1.6 R\textsubscript{E} outward motion.

\begin{figure}
\begin{centering}
\includegraphics{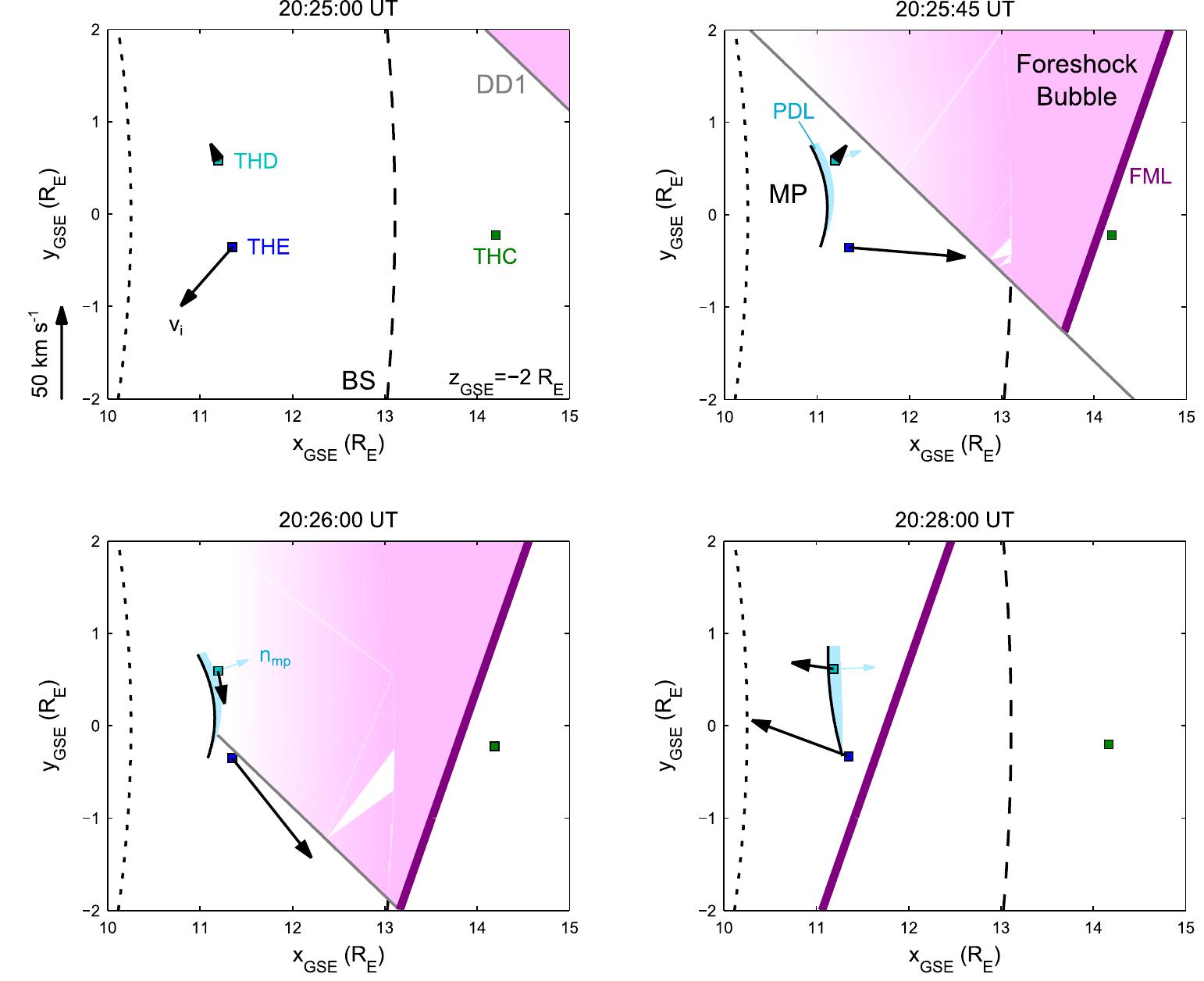}
\par\end{centering}

\caption{Schematics of the magentosheath/magnetopause response in the GSE z=-2~R\textsubscript{E}
plane at different times. Projections of the spacecraft positions
are shown as the coloured squares. The \citet{farris91} bow shock
and the \citet{shue98} magnetopause models are displayed as the dashed
and dotted lines respectively, whereas the observed magnetopause (with
normal $\mathbf{n}_{mp}$) is indicated by the solid black line along
with the plasma depletion layer (light blue). The discontinuity DD1
(grey) and fast mode layer (magenta) are also shown. The observed
ion velocities are given by the black arrows.\label{fig:msh-schematic}}

\end{figure}

Figure \ref{fig:msh-schematic} shows schematics of the magnetosheath
and magnetopause response to the FB at different times in the GSE
z=-2~R\textsubscript{E} plane. We indicate the observed magnetopause
(black solid line) using the calculated normals (light blue arrows)
at THD, the estimated magnetopause speed and the relative timings
of the crossings at both spacecraft. It can be seen in the top right
and bottom left panels that as DD1 tracked across the bow shock in
the vicinity of THD and THE, the magnetosheath flow (particularly
at THE) was accelerated towards the location of the intersection of
the current sheet with the shock.

Following the outbound crossing of the magnetopause at around 20:28:30-20:30:00
UT, both THD and THE again observed enhanced magnetosheath flows of
similar magnitude and duration to those at 20:26 in the vicinity of
DD1. The directions of these flows, however, was different being predominantly
directed anti-sunwards. The enhanced flows coincided with the observations
of DD2 at the spacecraft and the total pressure in the magnetosheath
exhibited an enhancement of $\sim$0.7~nPa. The timing of this pressure
enhancement at the two magnetosheath spacecraft was consistent with
the arrival of the FML at the spacecraft, again assuming unchanged
orientation and speed \citep[c.f.][]{sibeck03}, as shown in the bottom
right panel of Figure \ref{fig:msh-schematic}. Thus the rapid expansion
of the FB as it convected to the bow shock means that the $\sim$45~s
duration observations at THC upstream of the bow shock, resulted in
the observed $\sim$5~min duration impacts in the magnetosheath.

\subsection{Ground-magnetometer}

\begin{figure}
\begin{centering}
\includegraphics{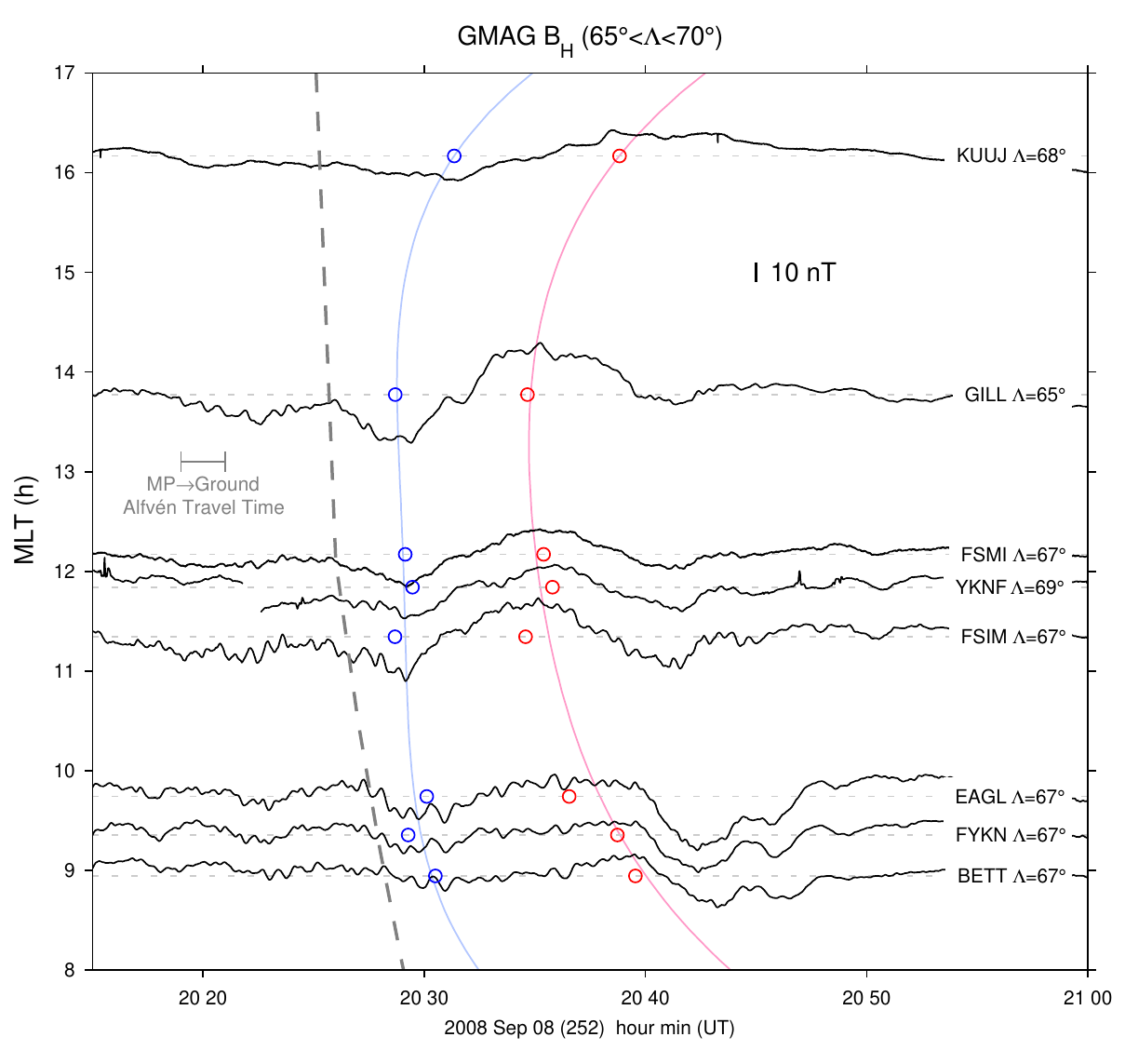}
\par\end{centering}

\caption{Stacked plots of the H-component of the magnetic field from ground-magnetometer
stations ordered by the average magnetic local time (MLT), where the
1~hour linear trend has been removed. Station names and geomagnetic
latitudes are noted. The minima (blue) and maxima (red) of the magnetic
impulse event (from 2~min smoothed data) at each station are indicated
by circles. The grey dashed line shows the intersection of DD1 with
the \citet{shue98} model magnetopause in the GSE z=-2~R\textsubscript{E}
plane and the Alfv\'{e}n travel time from the subsolar magnetopause
to the ground is also indicated.\label{fig:GMAG}}

\end{figure}

In addition to the spacecraft observations, we use ground magnetometer
(GMAG) data across North America taken from the THEMIS, GIMA, CARISMA
and CANMON networks in order to assess the scale of the magnetospheric
impact of the FB. Figure \ref{fig:GMAG} displays the H-component
(horizontal component towards mean magnetic north) from all available
GMAG stations at magnetic latitudes 65\textdegree{}$<\Lambda<$70\textdegree{},
ordered by magnetic local time (MLT). The one hour linear trend has
been removed from each of these to highlight the variations. The data
reveals a decrease followed by an increase at all stations, consistent
with an outward followed by inward motion of the magnetopause. We
estimate that the Alfv\'{e}n travel time from the subsolar magnetopause
to the FSMI GMAG (near noon MLT) was $\sim$2~min using the T96 magnetic
field model \citep{t95,t96}, in excellent agreement with the time
delay between the spacecraft observations of the boundary and the
extremum of the negative deflection observed on the ground.

The magnetic deflections were first observed and were strongest at
GILL at around 14:00 MLT. This corresponds to a bow shock $\theta_{Bn}\sim$40\textdegree{}
(see Figure \ref{fig:event-schematic}) i.e. approximately near the
edge of the ion foreshock \citep{lerussell92}. We associate this
with the arrival of the FB at the bow shock/magnetopause. All the
other stations observed deflections in quick sucession. We plot the
location of the intersection of DD1 with the \citet{shue98} magnetopause
model as a function of time as the grey dashed line in Figure \ref{fig:GMAG}.
This shows good agreement with the onset of the negative deflections
at the GMAG stations at the subsolar and morning sectors. We also
highlight the propagation of this magnetic impulse event through the
circles in Figure \ref{fig:GMAG}, which indicate minima (blue) and
maxima (red) from 2~min smoothed time series. Again these show that
both the outward and inward disturbances of the magnetopause originated
around 14:00 MLT and they subsequently propagated down both magnetopause
flanks. While the timings in the morning sector are similar to that
of the current sheet's transit across the dayside magnetopause, the
later observation in the afternoon sector at KUUJ cannot be explained
in this way. This could either be due to the initial arrival of the
FB at around 14:00 MLT launching a rarefaction wave down this flank
similar to the magnetospheric response to the arrival of solar wind
pressure pulses \citep{sibeck90} or perhaps due to the spatial structure
of the FB (see e.g Figure \ref{fig:FB-cartoon} or \citet{omidi10}).

By comparing the magnetic deflections observed by all stations at
a given time, it is possible to estimate the instantaneous scale size
of the magnetopause disturbance. For instance at around 20:30 UT $B_{H}$
was reduced at all GMAG stations i.e. there was a large scale outward
disturbance of the magnetopause from its nominal position spanning
at least 7~hours of magnetic local time. Similarly at around 20:38
UT $B_{H}$ was enhanced at all stations i.e. a large scale inward
disturbance of the boundary. Using the \citet{shue98} model magnetopause,
we find that these correspond to magnetopause disturbances spanning
at least 21.5~R\textsubscript{E} transerve to the Sun-Earth line.

\section{Discussion}

We have presented observations of a foreshock transient upstream of
Earth's bow shock along with simultaneous observations in the magnetosheath
and on the ground showing its magnetospheric impacts. We concluded
that the transient was unlikely to be an HFA, but that the observations
were in agreement with the understood properties of FBs. Here we piece
together all the observations, comparing the responses to those previously
simulated \citep{omidi10} as well as contrasting them with the known
effects of HFAs \citep[e.g.][]{sibeck99}.

\citet{omidi10} predicted that on arrival of an FB at the bow shock,
the reduced total pressure of the FB core should cause a reversal
of the magnetosheath flow back towards this low pressure region i.e.
sunwards. \citet{archer14} observationally showed that the reduced
total pressure upstream of the bow shock due to a foreshock transient
(most likely an HFA) resulted in strong (thermal + magnetic) pressure
gradients in the magnetosheath. They quantitatively demonstrated that
these pressure gradients directly accelerated the magnetosheath plasma,
causing fast sunward flows followed by outward magnetopause motion.
Here we have shown that magnetosheath plasma was accelerated towards
the intersection of the discontinuity with the bow shock, which quickly
tracked across the bow shock. This point at the shock separates the
original magnetosheath plasma (downstream) and the foreshock bubble
plasma (upstream). Since the FB reduced the total pressure upstream
of the bow shock, a pressure gradient must have existed. It is expected
that this pressure gradient would accelerate magnetosheath plasma
towards the intersection point, as observed. Whilst we cannot measure
this pressure gradient due to a lack of spacecraft, we interpret the
acceleration of the magnetosheath plasma towards this point as being
directly driven by pressure forces \citep{archer14} due to the arrival
of the FB.

The large scale outward motion of the magnetopause reported here due
to the FB's reduced total pressure was also predicted in the \citet{omidi10}
simulations. This is in stark contrast to the known effects of HFAs
on the magnetopause. HFAs have dimensions across the bow shock of
up to $\sim$4~R\textsubscript{E} \citep{facsko09}, which result
in similarly sized outward bulging of the boundary that moves with
the HFA's slow transit across the shock \citep{sibeck99,archer14}.
Observational studies have identified these distortions from the determined
boundary normals, revealing large deflections ($\geq$30\textdegree{})
from those expected by magnetopause models. In contrast, here the
estimated boundary normals were close to the model ones, with at least
the outbound crossing consistent with them being equal. Furthermore,
the ascertained orientation of the inbound normal was inconsistent
with that expected from a localised bulge of the magnetopause moving
with the intersection of DD1 with the bow shock. The leading edge
of such a bulge should be deflected from the model in the -y and -z
GSE directions \citep[e.g.][]{archer14} given the orientation of
DD1 (see Figure \ref{fig:msh-schematic}), however the observed (small)
deflection was in the +y direction. Depending on the spacecraft location,
normals aligned with the model boundary could be observed near the
peak of the bulge due to an HFA. However, the GMAG data shows further
evidence against such localised bulging, given the magnetic deflections
instantaneously spanned the equivalent of 21.5~R\textsubscript{E}
across the magnetopause.

The propagation and evolution of the magnetic impulse events observed
by the GMAGs was also unlike that known for HFAs. These have been
shown to move in agreement with the motion of their respective current
sheets across the bow shock \citep{eastwood08,jacobsen09}, growing
in time. This would correspond to a dusk to dawn motion. The effects
of the FB were first observed at an MLT in agreement with the edge
of the foreshock. Theoretically this is where the FB would first arrive
at the bow shock (see Figure \ref{fig:FB-cartoon}). The magnetic
impulse event subsequently propagated tailward in both the dawn and
dusk sectors, with decreasing amplitude. While the propagation dawnwards
is in agreement DD1's transit across the magnetopause, the duskwards
propagation may be due to the FB's transverse structure (c.f. Figure
\ref{fig:FB-cartoon}, \citet{omidi10,omidi13} and \citet{karimabadi14})
or through waves launched in the magnetosphere at the arrival of the
transient.

\section{Conclusions}

We have presented a case study of the impacts of a foreshock transient
on the magnetopause. From spacecraft observations upstream of the
bow shock, we showed that the transient was largely inconsistent with
a Hot Flow Anomaly (HFA). Instead it did agree with the understood
properties of Foreshock Bubbles (FBs). The FB modified the upstream
pressure of the solar wind, reducing the total pressure in its core
and sheath regions. Due to these pressure variations the FB impinged
upon the magnetosphere resulting in acceleration of magnetosheath
plasma towards the intersection of the current sheet with the bow
shock and large scale outward magnetopause motion spanning simultaneously
at least 7 hours of magnetic local time, equivalent to 21.5~R\textsubscript{E}
transverse to the Sun-Earth line at the magnetopause. Therefore, we
have shown for the first time that FBs, unlike other transient foreshock
phenomena such as HFAs \citep[e.g.][]{sibeck99} or foreshock cavities
\citep[e.g.][]{turner11}, have large scale/global impacts upon the
magnetosphere, similar to those of solar wind pressure pulses \citep[e.g.][]{sibeck90}.
It is not clear, however, in this event to what degree the large scale
disturbance of the magnetopause is an effect of the FB's transverse
scale size, which are simulated to be comparable to the extent of
the quasi-parallel shock \citep{omidi10}, or the fast transit of
the current sheet across the bow shock. Multi-spacecraft studies of
further FB events and their magnetospheric impacts may help in this
regard.

Magnetospheric ULF waves in the Pc5 (2-7~mHz) range generated by
sudden changes in the solar wind dynamic pressure have recently been
shown to have effects upon radiation belt electrons through bounce
and drift resonances \citep{claudepierre13,mann13}. Foreshock transients,
including FBs, are also a known source of Pc5 waves \citep{hartinger13}.
By utilising the constellation of spacecraft currently in the near-Earth
space environment (including Cluster, THEMIS, GOES and Van Allen Probes)
it is now possible to investigate the role that foreshock transients,
such as FBs, play in terms of the radiation belts, which may form
the subject of future work. While only a few examples of FBs have
been reported in the literature, a statistical study into their occurrence
and properties under different solar wind conditions could illuminate
how often this recently discovered phenomenon occurs. Furthermore,
additional simulations would also provide insight into their formation
and large scale structure under different current sheet orientations.

\section*{Acknowledgements}

M. O. Archer would like to thank C. H. K. Chen and H. Hietala for
helpful discussions. This research at Imperial College London was
funded by STFC grants ST/I505713/1, ST/K001051/1 and ST/G00725X/1.
D. L. Turner is thankful for funding from NASA (THEMIS mission and
grant NNX14AC16G). We acknowledge NASA contract NAS5-02099 and V.
Angelopoulos for use of data from the THEMIS Mission, specifically
C. W. Carlson and J. P. McFadden for use of ESA data; D. Larson and
R. P. Lin for use of SST data; J. W. Bonnell and F. S. Mozer for use
of EFI data; and K. H. Glassmeier, U. Auster and W. Baumjohann for
the use of FGM data provided under the lead of the Technical University
of Braunschweig and with financial support through the German Ministry
for Economy and Technology and the German Center for Aviation and
Space (DLR) under contract 50 OC 0302. We acknowledge A. Szabo and
K. Ogilvie for WIND data; and N. Ness and D. McComas for ACE data.
For GMAG data we thank the Geophysical Institute Magnetometer Array
operated by the Geophysical Institute, University of Alaska; I.R.
Mann, D.K. Milling and the rest of the CARISMA team, operated by the
University of Alberta and funded by the Canadian Space Agency; The
Canadian Magnetic Observatory Network (CANMON) maintained and operated
by the Geological Survey of Canada; S. Mende, C. T. Russell and NSF
for support through grant AGS-1004814.

\appendix

\section{Suprathermal density derivation\label{sec:Suprathermal-flux-derivation}}

\begin{figure}
\begin{centering}
\includegraphics{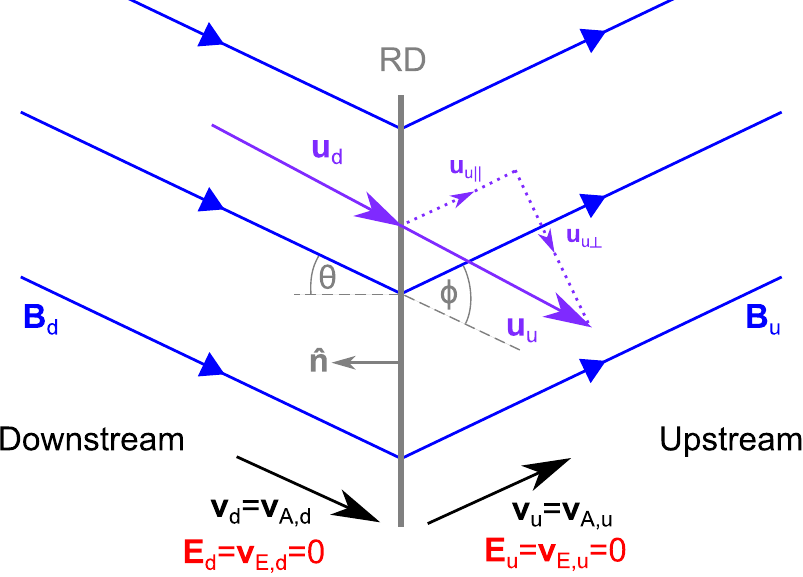}
\par\end{centering}

\caption{deHoffman-Teller rest frame of a rotational disconituity (RD, grey)
in the same format as Figure \ref{fig:FB-cartoon}.\label{fig:deHoffman-Teller-cartoon}}

\end{figure}

Here we derive the density of backstreaming suprathermal ions from
the foreshock on the upstream side of a rotational discontinuity in
its deHoffmann-Teller frame \citep{dehoffman50}. In Figure \ref{fig:deHoffman-Teller-cartoon}
we show an RD which rotates the tangential magnetic field by 180\textdegree{}
for simplicity (the angle between the upstream and downstream fields
$\phi=2\theta$ in this case, where 0\textdegree{}$<\theta<$90\textdegree{}
is the angle between the magnetic field and the RD normal $\hat{\mathbf{n}}$),
however we use the general case in our derivation whereby 0\textdegree{}$<\phi\leq2\theta$\textdegree{}.
Since the electric field in this frame is zero on both sides, particle
energies are conserved. Considering a single backstreaming ion with
velocity $\mathbf{u}_{d}$ on the downstream side, once it has crossed
the RD the ion will instantaneously have a velocity on the upstream
side $\mathbf{u}_{u}$ given by (assuming the thickness of the RD
is small relative to the upstream gyroradius):

\begin{align*}
\mathbf{u}_{u} & =\mathbf{u}_{d}\\
 & =u_{d}\mathbf{\hat{B}}_{d}\\
 & =u_{d}\cos\phi\mathbf{\hat{B}}_{u}+u_{d}\sin\phi\mathbf{\hat{u}}_{u\perp}
\end{align*}
i.e. the field-aligned guiding centre motion and a perpendicular component
which is the particle's gyrovelocity since there are no perpendicular
drifts in the deHoffmann-Teller frame. The incoming gyro-averaged
particle flux (on the downstream side of the RD) is

\begin{align*}
F_{in} & =n_{d}\mathbf{\overline{u}}_{d}\cdot\hat{\mathbf{n}}\\
 & =n_{d}u_{d}\cos\theta
\end{align*}
whereas the outgoing gyro-averaged flux (on the upstream side of the
RD) is
\begin{align*}
F_{out} & =n_{u}\mathbf{\overline{u}}_{u}\cdot\hat{\mathbf{n}}\\
 & =n_{u}u_{d}\cos\phi\mathbf{\hat{B}}_{u}\cdot\hat{\mathbf{n}}\\
 & =n_{u}u_{d}\cos\phi\cos\theta
\end{align*}
where $n$ is the number density of the particles under investigation
here. Conservation of particles across the RD requires $F_{in}=F_{out}$
thus

\begin{align*}
n_{d}u_{d}\cos\theta & =n_{u}u_{d}\cos\phi\cos\theta\\
n_{d} & =n_{u}\cos\phi
\end{align*}
i.e. the density of the suprathermal particles upstream of the RD
$n_{u}$ is greater than that incident $n_{d}$.

\section*{Bibliography}

\bibliographystyle{elsarticle-harv}
\addcontentsline{toc}{section}{\refname}\bibliography{foreshockbubble}

\end{document}